\documentstyle{article}

\input amssym.def
\input amssym.tex

\newcommand\newsection[1]{\refstepcounter{section}
\section*{} \vskip -1cm 
\noindent{\bf{\arabic{section}. {#1}}}
\message{\thesection. #1} 
\vskip.3cm \setcounter{equation}{0}
\renewcommand{\theequation}
{\arabic{section}.\arabic{equation}}}

\newcommand\nnewsection[1]{\vskip.5cm 
\noindent{\bf{#1}}\vskip.3cm 
\message{#1}
\setcounter{equation}{0}}

\def\boxit#1{\vbox{\hrule\hbox{\vrule\kern3pt
\vbox{\kern3pt#1\kern3pt}\kern3pt\vrule}\hrule}}

\def\sd{self-dual}
\def\asd{anti-self-dual}
\def\hc{hyper-complex}
\def\hk{hyper-k{\"a}hler}
\def\centreline{\centerline}
\def\cs{complex-structure}
\def\acs{almost-complex-structure}
\def\Rf{Ricci-flat}
\def\iqv{imaginary-quaternion-value}

\newlength{\extraspace}
\newlength{\extraspaces}
\newcommand{\be}{\begin{equation}
\addtolength{\abovedisplayskip}{\extraspaces}
\addtolength{\belowdisplayskip}{\extraspaces}
\addtolength{\abovedisplayshortskip}{\extraspace}
\addtolength{\belowdisplayshortskip}{\extraspace}}
\newcommand{\ee}{\end{equation}}

\newcommand{\ba}{\begin{eqnarray}
\addtolength{\abovedisplayskip}{\extraspaces}
\addtolength{\belowdisplayskip}{\extraspaces}
\addtolength{\abovedisplayshortskip}{\extraspace}
\addtolength{\belowdisplayshortskip}{\extraspace}}
\newcommand{\ea}{\end{eqnarray}}

\newcommand{\bann}{\begin{eqnarray*}
\addtolength{\abovedisplayskip}{\extraspaces}
\addtolength{\belowdisplayskip}{\extraspaces}
\addtolength{\abovedisplayshortskip}{\extraspace}
\addtolength{\belowdisplayshortskip}{\extraspace}}
\newcommand{\eann}{\end{eqnarray*}}

\newcommand{\form}[1]{{\Lambda}^{#1}}
\newcommand{\cmform}[1]{{\Lambda}_{c}^{#1}}

\def\M{M}
\def\TM{T{\M}}
\def\TSM{T^*{\M}}
\def\TCM{T_c {\M}}
\def\df{divergence-free}
\def\uh{ultra-hyperbolic}
\def\Uh{Ultra-hyperbolic}

\def\vpm{{\Bbb V}^{\pm}}
\def\vmp{{\Bbb V}^{\mp}}
\def\vp{{\Bbb V}^+}
\def\vm{{\Bbb V}^-}
\newcommand\vx[2]{{\Bbb V}^{#1}_{#2}}

\def\SLTC{{\rm SL}(2, {\Bbb C})}
\def\sltc{{\frak sl}(2, {\Bbb C})}

\begin{document}
\rightline{January 31, 1997}
\rightline{NCL96-TP1}
\rightline{gr-qc/9702001}
\vskip .5cm
\centreline{\bf{\Large Self-dual 
two-forms and divergence-free vector fields}} 
\vskip .5cm
\centreline{\bf{James D.E. Grant}}
\vskip .5cm
\centreline{Department of Physics, 
University of Newcastle,}
\centreline{Newcastle Upon Tyne, 
NE1 7RU, United Kingdom.}
\vskip .5cm
\begin{abstract}
\noindent Beginning with the self-dual two-forms approach to the 
Einstein equations, we show how, by choosing basis spinors 
which are proportional to solutions of the Dirac equation, 
we may rewrite the vacuum Einstein equations in terms of a 
set of divergence-free vector fields, which obey a 
particular set of chiral equations. Upon imposing the 
Jacobi identity upon these vector fields, we reproduce a 
previous formulation of the Einstein equations linked 
with a generalisation of the Yang-Mills equations for 
a constant connection on flat space. This formulation 
suggests the investigation of some new aspects of the 
self-dual two-forms approach. In the case of real 
Riemannian metrics, these vector fields have a natural 
interpretation in terms of the torsion of the natural 
almost-complex-structure on the projective spin-bundle.
\end{abstract}

\newsection{Introduction}

If one adopts a null tetrad approach 
to Lorentzian geometry in four 
dimensions, one quickly finds that, 
due to the structure of the 
Lorentz group, the 
spin connection and curvature 
quantities that arise break up 
naturally into self-dual and anti-self-dual 
quantities, related to one another 
by complex conjugation \cite{PR}. 
Given that the complexified space 
of $2$-forms also has a direct sum 
decomposition into self-dual and anti-self-dual 
$2$-forms, and that the self-dual part 
of the spin connection defines a natural connection 
on the space of self-dual $2$-forms, 
it is perhaps not too surprising 
that one can recast the equations 
of Lorentzian geometry solely in 
the language of self-dual $2$-forms 
and the self-dual spin-connection 
\cite{CDJM,MF}, along with the reality 
condition that self-dual objects are 
the complex conjugate of their 
anti-self-dual counterparts. 

One can adopt a similar approach in the more general
context of complex-Riemannian geometry, with the usual
description in terms of the bundle of orthonormal frames
and the natural ${\rm SO} (4, {\Bbb C})$ connection being
replaced with the bundle of self-dual $2$-forms, and its
$\sltc$ connection. Real slices of Lorentzian,
real-Riemannian or {\uh} signature then arise
when suitable reality conditions are satisfied.

The reformulation of four-dimensional geometry in such
terms becomes of special interest when we consider metrics
which satisfy the Einstein condition
\[
r = \frac{s}{4}\, {\bf g},
\]
where $r$ and $s$ denote the Ricci tensor and scalar
curvature of the metric ${\bf g}$. This condition is
equivalent to the condition that the self-dual
spin-connection have self-dual curvature. 
Although this observation
has been the basis of extensive work on the Lorentzian 
Einstein equations, it also seems to be of interest
in the context of compact Riemannian 
Einstein manifolds. In this case the self-dual 
spin-connection is an self-dual
${\rm SU}(2)$ Yang-Mills field with second chern class
$c_2 = - \frac{1}{2} (\chi + 
\frac{3}{2} \tau )$. A large amount of 
information has accumulated concerning the 
moduli spaces of such fields on generic 
four-manifolds \cite{DK}, at least
for small values of $c_2$, although it is not clear 
at this point whether this information is
generally useful in the study 
of existence and uniqueness of 
Einstein metrics\footnote{I am 
grateful to Andrew Chamblin for
discussions on this point.}.

Another area in which this 
alternative description of the
Einstein condition is useful 
is in the context of
{\Rf} metrics with anti-self-dual 
Weyl tensor in which case 
the self-dual connection is 
flat. If we specialise to a simply
connected, real Riemannian 
manifold $\M$, a {\Rf}
anti-self-dual metric automatically 
defines a {\hk}
structure. The three K{\"a}hler 
forms associated with this
{\hk} structure 
can then be identified with the
self-dual $2$-forms of 
our formalism. Alternatively
\cite{MN}, the integrability condition 
for the three {\cs}s can be recast 
as the self-dual Yang-Mills
equations for a constant 
connection on flat space, with the
connection taking values 
in {\em Lsdiff}$\M$, the algebra
of {\df} vector 
fields on $\M$. Building upon
this result, it was shown 
that one could reformulate the
full vacuum Einstein equations 
in terms of such a set of
{\df} vector 
fields \cite{G}. Unlike 
the self-dual
$2$-forms approach to the 
Einstein equations mentioned
above, this reformulation is 
symmetrical between the
self-dual and anti-self-dual 
parts of the gravitational field and,
although the final equations 
bear some resemblance to the
full Yang-Mills equation 
for a constant connection on flat
space, they also included 
interaction terms between {\sd} and 
{\asd} parts of the gravitational field. 

Our aim here is to relate the 
self-dual $2$-forms approach
to the Einstein equations 
with the {\df} vector
field approach. 
We begin by reviewing 
the description of four-dimensional 
Lorentzian geometry in terms 
of two component spinors, null tetrads 
and self-dual $2$-forms. We then show that 
by choosing appropriate
spinor bases and carrying 
out a related conformal
transformation, we can 
describe any metric in terms of a
set of {\df} vector 
fields, which obey a set 
of equations which are 
explicitly chiral in nature. 
Reversing the argument, we see 
that the vector field approach 
suggests the investigation of a 
particular set of divergence-free 
vector fields which occur naturally 
in the self-dual $2$-forms approach. 
These divergence-free vector fields 
essentially carry the information of 
the {\sd} spin connection and, 
for Riemannian metrics, 
can be intepreted in terms of the 
torsion of an {\acs} on the tangent 
space of our manifold. 
This {\acs} is the projection 
to $\TM$ of the horizontal part 
of the natural {\acs} on the projective 
spin bundle which arises when we 
consider metrics with {\asd} Weyl tensor. 
The Einstein condition tells us that 
the Riemann tensor evaluated on any 
{\asd} null bivector in the 
complexification of the tangent space 
must commute with the {\acs}. 

Most of the reality conditions used 
in this paper are those necessary for 
the description of metrics of Lorentzian 
signature. Since, however, 
Section~\ref{sec:conn} is concerned with 
Riemannian metrics, we have included an 
appendix devoted to a general discussion 
of the reality conditions for metrics of 
Riemannian and {\uh} signature.

\newsection{Self-dual two-forms}
\label{sec:notn}

We assume we are on a smooth, 
oriented, real four-manifold
$\M$ with a pseudo-Riemannian 
metric, ${\bf g}$, of
Lorentzian signature $(+---)$. 
The manifold $(\M, {\bf
g})$ comes naturally equipped 
with the bundle of exterior
p-forms, $\form{p}$. The Hodge map is the
unique vector bundle isomorphism
\[
* : \form{k} \rightarrow \form{4-k},
\]
defined by 
\be
\alpha \wedge *\beta =
{\bf g} (\alpha, \beta) \nu, \qquad 
\alpha, \beta \in \form{k}, 
\label{hodge}
\ee
where ${\bf g}(\alpha, \beta)$ is 
the product on $\form{k}$ induced 
by the metric, and $\nu$ is the 
volume form defining the orientation 
\cite{AHS,Be}. In the particular case 
of four dimensional Lorentzian spacetimes, 
the Hodge map acts as an 
endomorphism of ${\Lambda}_x^2$ with 
$*^2 = - 1$, so the space ${\Lambda}_x^2$ 
has a natural {\cs} \cite{Be,Br}. Therefore 
the complexified bundle of $2$-forms 
$\cmform{2} = \form{2} \otimes {\Bbb C}$ 
decomposes as $\cmform{2} = \cmform{2+} 
\oplus \cmform{2-}$ where $\cmform{2\pm}$ 
are the bundles of self-dual 
and anti-self-dual $2$-forms:
\be
\cmform{2\pm} = \{ \lambda \in \cmform{2}: 
{}^* \lambda = \pm i \lambda \}.
\label{sdasdtfs}
\ee
Due to the use of the complexification of
$\form{2}$ in the decomposition, complex conjugation
defines isomorphisms $\cmform{2\pm} \cong 
{\overline{\cmform{2\mp}}}$. Note that 
the spaces $\cmform{2\pm}$ are orthogonal 
with respect to the product ${\bf g}$ when it is 
extended, by linearity, to $\cmform{2}$.

This decomposition of $\cmform{2}$ has a straightforward
interpretation in terms of Lie algebras. Since our metric
has Lorentzian signature, the orthonormal frame bundle has
structure group ${\rm SO}(1, 3)$. As an ${\rm SO}(1, 3)$
module, $\form{2}$ is isomorphic to the Lie algebra
$\frak{so}(1, 3)$ and, although the algebra $\frak{so}(1, 3)$
is simple, its complexification $\frak{so}(1, 3) \otimes
{\Bbb C}$ decomposes as $\sltc \oplus \sltc$. 
As in the Riemannian case \cite{AHS,Be}, we now
use this isomorphism at the group level, where ${\rm Spin}
(1, 3) \otimes {\Bbb C} \cong \SLTC \times \SLTC$, to
introduce the $2$-dimensional complex vector bundles
$\vpm$ of self-dual and anti-self-dual spinors. Since these are
$\SLTC$ bundles, there naturally come equipped with
symplectic forms ${\epsilon}_{\pm}$, which can be used to
define isomorphisms $\vpm \cong (\vpm)^*$ between $\vpm$
and their duals $(\vpm)^*$. As in the decomposition
of $\cmform{2}$ complex conjugation defines 
isomorphisms $\vpm \cong {\overline{\vmp}}$, 
denoted by $\pi \mapsto {\overline{\pi}}$.

Given the basic spin-bundles $\vpm$, general 
spin-bundles are constructed 
by taking appropriate symmetric products, 
with a field $\phi \in \Gamma 
(S^m \vm \otimes S^n \vp)$ transforming under the 
irreducible representation $(m,n)$ of 
$\SLTC \times \SLTC$. Extending 
complex conjugation to the 
higher bundles, the bundles 
$S^m \vm \otimes S^m \vp$ inherit 
real structures for each positive 
integer $m$. In particular, from the 
vector representation of ${\rm SO}(1, 3)$, 
we deduce that the complexified 
tangent bundle of a Lorentzian 
four-manifold is isomorphic to 
the product of spin-spaces
\[
\TCM \equiv \TM \otimes {\Bbb C} \cong \vm \otimes \vp,
\]
with the real tangent bundle $\TM$ 
corresponding to products of spinors 
invariant under the 
real structure on $\vm \otimes \vp$. 
Using this isomorphism, we can translate a given
complexified tensor field into a section of a given spinor
bundle, and then reduce it to irreducible spinor parts by
use of the $\SLTC$ invariant forms ${\epsilon}_{\pm}$, and
the decomposition of the tensor product
\[
S^m \vpm \otimes S^n \vpm \cong 
{\oplus}_{k = 0}^{min(m,
n)} S^{m+n-k} \vpm. 
\]
For example, the tensor product 
of $\cmform{1}$ with itself
decomposes as
\[
{\otimes}^2 \cmform{1} \cong
\cmform{2+} \oplus \cmform{2-} 
\oplus S_0^2\cmform{1} \oplus {\Bbb C}{\bf g},
\]
where $S_0^2\cmform{1}$ denotes the 
trace-free symmetric tensors (the trace
being defined with the metric ${\bf g}$). 
We can then identify
the complexified metric in the form 
\be
{\bf g}
\cong {\epsilon}_+ \otimes {\epsilon}_-.
\label{metdec}
\ee
Any complex null vector ${\bf v} 
\in (\TCM)_x$ can be written
\[
{\bf v} \cong \pi \otimes \lambda,\qquad
{\rm where} \qquad \pi \in (\vm)_x, \qquad 
\lambda \in (\vp)_x,
\]
and a real null vector ${\bf v} 
\in (\TM)_x$ may be expressed as
\[
{\bf v} \cong \pi \otimes {\overline{\pi}},\qquad
{\rm where} \qquad \pi \in (\vm)_x, \qquad 
{\overline{\pi}} \in (\vp)_x.
\]

In particular, if we introduce a local basis 
$\{ {\epsilon}_A : A = 0, 1\}$ for $\vm$, 
and the dual basis $\{ {\epsilon}^A \}$ 
for $(\vm)^*$, we can, without loss of
generality, assume that the bases are 
orthonormal, in the sense that the
components of ${\epsilon}_-$ with 
respect to $\{ {\epsilon}_A \}$ are 
\be
{\epsilon}_{AB} \equiv {\epsilon}_- 
( {\epsilon}_A, {\epsilon}_B ) 
= \left(\matrix
{0&1\cr
-1&0\cr}\right).
\label{epdown}
\ee
The dual bundle $(\vpm)^*$ also inherits 
a symplectic structure ${\epsilon}_{\pm}^*$ 
which, relative to the basis 
$\{ {\epsilon}^A \}$, has components 
\be
{\epsilon}^{AB} \equiv
{\epsilon}_-^* ( {\epsilon}^A, {\epsilon}^B) = 
\left(\matrix
{0&1\cr
-1&0\cr}\right).
\label{epup}
\ee
We can raise and lower spinor indices using the
components of the symplectic forms 
${\epsilon}_{\pm}$ according
to the standard conventions
\[
{\lambda}^A = {\epsilon}^{AB} {\lambda}_B, \qquad
{\lambda}_A = {\lambda}^B {\epsilon}_{BA}.
\]
Similar orthonormal bases, 
denoted $\{ {\epsilon}_{A^\prime} : 
A^{\prime} = 0^{\prime}, 1^{\prime} \}$ 
and $\{ {\epsilon}^{A^\prime} \}$ can be
introduced for for $\vp$, and $(\vp)^*$. 
We can, without loos of generality, 
choose the bases ${\epsilon}_A$ and 
${\epsilon}_{A^\prime}$ to be 
related by complex conjugation with
\be
{\overline{{\epsilon}_A}} 
= {\epsilon}_{A^\prime},
\qquad
{\overline{{\epsilon}_{A^\prime}}} 
= {\epsilon}_A.
\label{lreal}
\ee

If we denote these local bases for $\vpm$
by ${\epsilon}_A \equiv (o, \iota)$ and 
${\epsilon}_{A^{\prime}} \equiv ({o}^{\prime}, 
{\iota}^{\prime})$, then we can 
define the null basis for $\TCM$
\be
{\bf e}_1 \cong o 
\otimes o^{\prime}, \qquad
{\bf e}_2 \cong \iota 
\otimes {\iota}^{\prime}, \qquad
{\bf e}_3 \cong o 
\otimes {\iota}^{\prime}, \qquad
{\bf e}_4 \cong \iota 
\otimes o^{\prime}.
\label{NPvecs}
\ee
The components of the metric (\ref{metdec}) 
with respect to this basis is given by the matrix
\[
{\eta}_{ab} \equiv
{\bf g} ( {\bf e}_a, {\bf e}_b) = 
\left(\matrix
{0&1&0&0\cr
1&0&0&0\cr
0&0&0&-1\cr
0&0&-1&0\cr}\right).
\]
Therefore, in terms of the dual basis 
$\{ {\bf \epsilon}^a \}$ for $\cmform{1}$, 
the metric may be expressed as
\be
{\bf g} = 
{\bf \epsilon}^1 \otimes {\bf \epsilon}^2
+{\bf \epsilon}^2 \otimes {\bf \epsilon}^1
-{\bf \epsilon}^3 \otimes {\bf \epsilon}^4
-{\bf \epsilon}^4 \otimes {\bf \epsilon}^3.
\label{NPmet}
\ee
Similarly, the inverse metric\footnote{Throughout, 
we will make use of the standard musical 
isomorphisms ${}^{\flat}: \TM \rightarrow \TSM$ 
and ${}^{\sharp}: \TSM \rightarrow \TM$, defined 
using the metric ${\bf g}$, and their extensions 
to arbitrary tensor bundles} ${\bf g}^{\sharp} 
\in \Gamma( S^2 (\TCM))$ takes the form
\be
{\bf g}^{\sharp} = 
{\bf e}_1 \otimes {\bf e}_2
+{\bf e}_2 \otimes {\bf e}_1
-{\bf e}_3 \otimes {\bf e}_4
-{\bf e}_4 \otimes {\bf e}_3.
\label{NPinvmet}
\ee
Given the complex conjugation laws 
(\ref{lreal}) for the spinor bases, 
we deduce that the vector fields 
$\{ {\bf e}_i \}$ obey the reality relations
\[
{\overline{{\bf e}_1}} = {\bf e}_1,
\qquad
{\overline{{\bf e}_2}} = {\bf e}_2,
\qquad
{\overline{{\bf e}_3}} = {\bf e}_4,
\qquad
{\overline{{\bf e}_4}} = {\bf e}_3,
\]
and the real tangent space takes the form
\[
\TM = {\rm Span}_{\Bbb R} 
\left( 
{\bf e}_1, {\bf e}_2, {\bf e}_3 + {\bf e}_4, 
i \left( {\bf e}_3 - {\bf e}_4 \right) 
\right).
\]
The complex metric (\ref{NPmet}) 
then indeed restricts to a real metric 
of Lorentzian signature on the real 
tangent space.

If we consider the Riemann 
curvature tensor of the metric
${\bf g}$, then, using the 
metric, this may be viewed as
self-adjoint map ${\cal R}: 
\form{2} \to \form{2}$ given
by 
\[
{\cal R} ({\bf \epsilon}^a 
\wedge {\bf \epsilon}^b) =
\frac{1}{2} R^{ab}{}_{cd} 
{\bf \epsilon}^c 
\wedge {\bf \epsilon}^d,
\]
where $\{ {\bf \epsilon}^a \}$ 
is a local orthonormal 
basis for $\form{1}$. In terms of 
the decomposition $\cmform{2} = 
\cmform{2+} \oplus \cmform{2-}$, 
${\cal R}$ can be put in
block form 
\be
\left(\matrix{
{}^+ W + \frac{s}{12}&{\Phi} \cr
{\Phi}^*&{}^- W + \frac{s}{12} \cr}
\right),
\label{blockcurv}
\ee
relative to orthonormal bases 
$\{ {}^{\pm} {\bf \lambda}^i
\}$ for $\cmform{2\pm}$. Using 
the isomorphism between tensors 
and spinors introduced above, 
${}^+ W$ and ${}^- W$ correspond
to the $S^4 \vp$ and $S^4 \vm$ 
parts of the Riemann tensor,
which can be identified with the 
self-dual and anti-self-dual parts
of the Weyl tensor respectively. 
${\Phi}$ corresponds to the 
$S^2 \vm \otimes S^2 \vp$ part 
of the curvature, which can be 
identified with the trace-free 
part of the Ricci tensor:
\[
{\Phi} = r - \frac{s}{4} {\bf g},
\]
where $s = {\rm tr} \ r$ denotes the 
scalar curvature. In Lorentzian 
signature, we may choose
the bases $\{ {}^{\pm} {\bf \lambda}^i \}$ 
to be complex conjugate to one another, 
in which case ${}^+ W$ and ${}^- W$ are
complex conjugates, ${\Phi}$ viewed as 
a $3 \times 3$ matrix is Hermitian, 
and the scalar curvature $s$ is real.

\vskip .4cm

In precisely four dimensions, there is an alternative
description of conformal geometry. Suppose we introduce a
set of three linearly-independent complex $2$-forms $\{
{\bf \Sigma}^i : i = 1, 2, 3 \}$ on a 
real four-manifold $\M$. These $2$-forms
will be self-dual with respect 
to a unique conformal class of
metrics on $\M$. If these $2$-forms
are orthogonal to their complex conjugates, with
\be 
{\bf \Sigma}^i \wedge {\overline{\bf \Sigma}}^j
= 0,\qquad i, j = 1, 2, 3, 
\label{lorreal}
\ee
then the conformal structure is of Lorentzian signature.
These complex $2$-forms may be combined into a single
$\sltc$ valued two-form: 
\[
{\bf \Sigma} = - \frac{i}{2} {\bf \Sigma}^i \tau_i,
\]
where $\tau_i$ are the Pauli matrices. 
In the same way that a frame for $\TM$ 
defines an isomorphism between $(\TM)_x$ 
and ${\Bbb R}^4$, the form ${\bf \Sigma}$ 
defines an isomorphism between 
$(\cmform{2+})_x$ and ${\Bbb C}^3$. Given 
${\bf \Sigma}$, we may define the unique 
$\sltc$-valued connection ${\bf \gamma}$ 
on the vector bundle $\cmform{2+}$ by the condition
\be
d {\Sigma} + 
\left[ \gamma , \Sigma \right] = 0,
\label{notorsion}
\ee
and the associated curvature
\be
R = d {\gamma} + \frac{1}{2} 
\left[ \gamma , \gamma \right].
\label{curvature}
\ee

Given that the $2$-forms ${\bf \Sigma}^i$ define an
isomorphism $(\cmform{2+})_x \cong {\Bbb C}^3$,
we can, by means of a GL$(3, {\Bbb C})$ transformation,
choose the ${\bf \Sigma}^i$ to obey the orthonormality
condition
\be 
{\bf \Sigma}^i \wedge {\bf \Sigma}^j =
i {\delta}^{ij} {\bf \nu}, 
\label{orthso3c}
\ee
with ${\bf \nu}$ a real volume element 
on $\M$. Similarly, the complex 
conjugate basis obey the relation
\[
{\bf{\overline\Sigma}}^i \wedge 
{\bf{\overline\Sigma}}^j =
- i {\delta}^{ij} {\bf \nu}, 
\]
along with the condition (\ref{lorreal}). 

In $\sltc$ language, since we can identify the adjoint
representation space of $\sltc$ with $S^2 \vp$, we can
represent the form ${\bf \Sigma} \in \Gamma ( \cmform{2}
\otimes \sltc )$ by its components 
${\bf \Sigma}_{A^{\prime}}{}^{B^{\prime}}$
with respect to the bases $\{ {\epsilon}_{A^{\prime}} \}$ and $\{
{\epsilon}^{A^{\prime}} \}$ for 
$\vp$ and $(\vp)^*$ introduced above,
with the condition: 
\[ 
{\bf \Sigma}_{A^{\prime}}{}^{A^{\prime}} = 0
\qquad
\Rightarrow 
\qquad {\bf \Sigma}^{{A^{\prime}}{B^{\prime}}} 
= {\bf \Sigma}^{{B^{\prime}}{A^{\prime}}}. 
\]
The orthonormality condition 
(\ref{orthso3c}) then becomes
\be
{\Sigma}_{{A^{\prime}}{B^{\prime}}} 
\wedge 
{\Sigma}^{{C^{\prime}}{D^{\prime}}} = 
i {\epsilon}_{({A^{\prime}}}{}^{C^{\prime}}
{\epsilon}_{{B^{\prime}})}{}^{D^{\prime}} {\bf \nu}.
\label{ort}
\ee
This means \cite{CDJM,MF} 
that there exists a basis, 
$\{ {\bf \epsilon}^a \cong 
{\bf \epsilon}^{AA^{\prime}} \}$, 
of $\cmform{1}$, unique up to an 
$\sltc$ rotation of the basis $\{
{\epsilon}_A \}$, 
with the property that 
\[
{\Sigma}_{{A^{\prime}}{B^{\prime}}} = \frac{1}{2} \,
{\epsilon}_{AB}\,
{\bf \epsilon}^{AA^{\prime}} \,\wedge\,
{\bf \epsilon}^{BB^{\prime}}.
\]
In terms of this basis, 
the metric may be written
\be
{\bf g} = {\epsilon}_{AB}\,
{\epsilon}_{A^{\prime}B^{\prime}}\,
{\bf \epsilon}^{AA^{\prime}} 
\otimes {\bf \epsilon}^{BB^{\prime}}. 
\label{metric}
\ee
We can also construct a basis for 
the space $\cmform{2-}$ of anti-self-dual 
$2$-forms given by
\[
{\Sigma}^{A^{\prime}B^{\prime}} = 
\frac{1}{2} {\epsilon}_{A^{\prime}B^{\prime}}\,
{\bf \epsilon}^{AA^{\prime}} \wedge 
{\bf \epsilon}^{BB^{\prime}},
\]
which are orthogonal to the 
self-dual $2$-forms 
${\Sigma}_{{A^{\prime}}{B^{\prime}}}$ in the sense that 
\be
{\Sigma}^{AB} \wedge 
{\Sigma}^{A^{\prime}B^{\prime}} = 0.
\label{sdasdorth}
\ee

When we identify ${\Sigma}$ in this way 
with the metric and tetrad, the 
connection ${\bf \gamma}$ of equation 
(\ref{notorsion}) becomes 
the self-dual part of 
the standard spin-connection, 
${\Gamma}^a {}_b$, defined by 
\be
d {\bf \epsilon}^a +
{\Gamma}^a {}_b \,\wedge\,{\bf \epsilon}^b = 0,
\qquad 
{\Gamma}^a {}_b =
{\Gamma}_c {}^a {}_b \, {\bf \epsilon}^c.
\label{spinconn}
\ee
The curvature $R$ in (\ref{curvature}) can 
then be identified with the elf-dual part 
of the Riemann curvature.
In the notation of equation 
(\ref{blockcurv}) we therefore have
\be 
R_{A^{\prime}B^{\prime}} = 
{}^+ W_{A^{\prime}B^{\prime}C^{\prime}D^{\prime}}
{\Sigma}^{C^{\prime}D^{\prime}} +
{\Phi}_{ABA^{\prime}B^{\prime}} {\Sigma}^{AB}
+ \frac{s}{12} {\Sigma}_{A^{\prime}B^{\prime}}. 
\label{curvdecomp}
\ee
The condition that the metric 
${\bf g}$ be Einstein is that 
${\Phi} = 0$, which, from (\ref{curvdecomp}), 
we see is equivalent to the condition that 
the self-dual spin-connection has
self-dual curvature \cite{CDJM,AHS,Br}:
\be 
{}^* R = R.
\label{sdcurv} 
\ee
The Einstein equations are fully 
characterised by equations 
(\ref{notorsion}), (\ref{ort}) and 
(\ref{sdcurv}). In particular, this 
interpretation of the Einstein 
condition only involves the frame 
and connection for $\cmform{2+}$ 
and there is no dependence 
(at least explicitly) upon the 
properties of the anti-self-dual part 
of the gravitational field.
 
We now introduce a dual basis $\{ {\bf e}_a \}$ 
for $\TCM$, where
\[
< {\bf \epsilon}^a, {\bf e}_b > 
= {\delta}^a_b.
\]
When acting as a differential operator, 
we will denote ${\bf e}_a$ 
by ${\nabla}_a$. We define the
commutator coefficients, $C_{ab}{}^c$, of the vector
fields ${\bf e}_a$ by the relation
\[
\left[ {\bf e}_a, {\bf e}_b \right] 
= C_{ab}{}^c {\bf e}_c,
\]
so that
\[
C_{ab}{}^c = 
< \left[ {\bf e}_a, {\bf e}_b \right], 
{\bf \epsilon}^c >.
\]

If we decompose the spin-connection 
${\Gamma}^a {}_b$ and the commutator 
coefficients $C_{ab}{}^c$ into spinor 
terms according to the formulae
\ba
{\Gamma}_a {}^b {}_c &\cong&
{\epsilon}_{B^{\prime}} {}^{C^{\prime}}
{\gamma}_{AA^{\prime}B} {}^C
+ 
{\epsilon}_B {}^C
{\gamma}_{AA^{\prime}B^{\prime}}
{}^{C^{\prime}},
\label{gamspin}
\\
C_{ab} {}^c &\cong& 
{\epsilon}_{A^{\prime}B^{\prime}}
C_{AB} {}^{CC^{\prime}}
+
{\epsilon}_{AB}
C_{A^{\prime}B^{\prime}} 
{}^{CC^{\prime}},
\label{cspin}
\ea
and use the standard relationship 
between the spin-connection and 
the commutator coefficients for 
a pseudo-orthonormal basis
\be
{\Gamma}_{abc} = \frac{1}{2}
\left[ 
C_{acb} - C_{abc} - 
C_{cba}
\right],
\label{gamcom}
\ee
then we find that
\ba
&C_{A^{\prime}B^{\prime}} {}^{CC^{\prime}}
= 
{\epsilon}_{(A^{\prime}} {}^{C^{\prime}}
{\gamma}_{B^{\prime})}
{}^{DC}
-
{\gamma}^C {}_{(A^{\prime}B^{\prime})}
{}^{C^{\prime}},&
\label{cgam}
\\
&{\gamma}_{AA^{\prime}B^{\prime}C^{\prime}} 
=
- \frac{1}{2} \left[
C_{A^{\prime}(B^{\prime}C^{\prime})A} + 
C_{B^{\prime}C^{\prime}A^{\prime}A} +
C_{AD}{}^{D}
{}_{(B^{\prime}} {\epsilon}_{C^{\prime})A^{\prime}}
\right].&
\label{gamc}
\ea

\newsection{Divergence-free vector fields}
\label{sec:weyl}

Up to now, our discussion has been of completely 
general tetrads and metrics. 
It is known, however, that if 
we consider {\Rf} metrics with {\asd} Weyl 
tensor, it is often advantageous to partially 
fix the internal frame ${\bf e}_i$ for $\TM$, 
and perform a related 
conformal transformation \cite{MN}. 
More precisely \cite{Wa}, 
if we have a conformal structure 
which is self-dual, with ${}^+ W = 0$, then it is 
possible to find a representative metric within 
the conformal class where the inverse metric 
takes the form 
\[
{\bf{\hat g}}^{\sharp} = {\eta}^{ab} 
{\bf{\hat e}}_a \otimes {\bf{\hat e}}_b, 
\]
for a fixed constant internal metric 
${\bf \eta}$, and where the basis vectors 
${\bf{\hat e}}_a$ obey the relation:
\[
\left[ {\bf{\hat e}}_a, {\bf{\hat e}}_b \right] = 
- \frac{1}{2} {\epsilon}_{ab}{}^{cd} 
\left[ {\bf{\hat e}}_c, {\bf{\hat e}}_d \right],
\]
where 
\[
{\epsilon}_{abcd} = 
\left\{ \begin{array}{ll}
1 & \mbox{if abcd an even permutation of 1234} \\
-1 & \mbox{if abcd an odd permutation of 1234} \\ 
0 & \mbox{otherwise}
\end{array}
\right.
\]
and indices are raised and lowered with the 
object ${\bf \eta}$ and its inverse. If we 
have a physical metric which is anti-self-dual and 
{\Rf}, we may also take the vectors 
${\bf{\hat e}}_a$ to be {\df} with respect 
to some volume element ${\bf{\hat \omega}}$, with
\[
{\cal L}_{{\bf {\hat e}}_a} {\bf {\hat \omega}} = 0.
\]
Defining the function $f$ by 
\[
{\bf {\hat \omega}}({\bf {\hat e}}_{1}, {\bf {\hat e}}_2, 
{\bf {\hat e}}_3, {\bf {\hat e}}_4) = f^2,
\]
then we obtain the physical, {\Rf} 
inverse metric by the conformal transformation
\[
{\bf g}^{\sharp} = f^2 {\bf{\hat g}}^{\sharp}.
\]

Given that such a conformal transformation 
is useful in the study of anti-self-dual {\Rf} 
metrics, it is natural to ask whether 
one can develop a similar approach to the full 
Einstein equations, without the anti-self-duality 
constraint \cite{G}. Since this {\df} condition is not 
preserved under a general internal rotation of 
our vector basis, however, it is necessary to 
study what, if any, restrictions such a choice 
of gauge places on the geometry. In particular, 
to make the transformation to {\df} vector fields, 
it is necessary to choose our vector basis so that 
the inverse metric is written 
\be
{\bf g}^{\sharp} = {\eta}^{ab} 
{\bf e}_a \otimes {\bf e}_b, 
\label{conmet}
\ee
where the vectors $\{ {\bf e}_a \}$ 
satisfy the condition
\be
C_{ab}{}^b = - {\nabla}_a (\log f),
\label{gauge}
\ee
for some function $f$ \cite{MN,G}. 
Our aim in this section is to show 
that we can always achieve this 
condition by an internal rotation 
of basis vectors, for any 
inverse metric ${\bf g}^{\sharp}$. 

Beginning with any real Lorentzian metric, we can, without
loss of generality, complexify the tangent space and
define a local basis $\{ {\bf e}_a \}$ for $\TCM$ where the
inverse metric takes the form (\ref{NPinvmet}). With the 
standard spin-connection defined by equation
(\ref{spinconn}), and the components of the
spin-connection and the commutator coefficients related by
equation (\ref{gamcom}), we see that the gauge condition
(\ref{gauge}) may be rewritten 
\[
{\Gamma}_b {}^b {}_a
= {\nabla}_a (\log f). 
\] 
This in turn implies that 
\be
{\rm div} \left( f^{-1} 
e_a \right) = 0,
\label{nabfree}
\ee
where, for an arbitrary 
vector field ${\bf v} \in 
\Gamma(\TCM)$, we have defined 
\[
{\rm div} \, {\bf v} = {\delta} {\bf v}^{\flat},
\]
where ${\delta}: \form{p+1} \rightarrow 
\form{p}$ is codifferentiation \cite{Be}.

As in the previous section, we can use the 
isomorphism $\TCM \cong \vm \otimes \vp$ to 
introduce bases $\{ {\epsilon}_A \}, 
\{ {\epsilon}_{A^{\prime}} \}$ 
for $\vm$ and $\vp$ in terms of which 
we can write the basis ${\bf e}_a$ as
\[
{\bf e}_a \cong {\epsilon}_A
\otimes {\epsilon}_{A^{\prime}}. 
\]
Defining spinor fields 
\be
{\alpha}_A = f^{-1/2} 
{\epsilon}_A,\qquad
{\alpha}_{A^{\prime}} = f^{-1/2} 
{\epsilon}_{A^{\prime}}
\label{adef}
\ee
then, from (\ref{nabfree}) 
we deduce that we require
\be
{\epsilon}_+ (D^- {\alpha}_A, 
{\alpha}_{A^{\prime}})
+ 
{\epsilon}_- 
( D^+ {\alpha}_{A^{\prime}}, 
{\alpha}_A) = 0,
\label{dfree}
\ee
where 
\[
D^{\pm}: \Gamma (\vpm) 
\to \Gamma (\vmp)
\] 
are the standard Dirac operators, 
defined by pulling back
the connection on the frame bundle 
to an $\sltc \times \sltc$ on the 
full spin-bundle, then projecting 
onto the separate $\sltc$ factors 
to give connections on $\vpm$. 
Assuming for the moment that we are given
solutions ${\alpha}_A, {\alpha}_{A^{\prime}}$ 
of equation (\ref{dfree}), we can 
straightforwardly reconstruct new 
spinor bases with the required gauge 
properties. Explicitly, we define 
\[ 
\chi = {\epsilon}_- ( {\alpha}_0, 
{\alpha}_1), \qquad
{\chi}^{\prime} = 
{\epsilon}_+ ({\alpha}_{0^{\prime}}, 
{\alpha}_{1^{\prime}}),
\]
then the spinor fields
\[
{\tilde \epsilon}_A = 
{\chi}^{-1/2} {\alpha}_A,\qquad
{\tilde \epsilon}_{A^{\prime}} = 
{{\chi}^{\prime}}^{-1/2} 
{\alpha}_{A^{\prime}},
\]
constitute normalised spinor bases. 
The vector fields
\[
{\tilde {\bf e}}_a \cong 
{\tilde \epsilon}_{A} \otimes 
{\tilde \epsilon}_{A^{\prime}}
\]
form a normalised basis for $\TCM$ which 
satisfy the condition (\ref{gauge}) 
with the function $f$ given by 
\[
f = ( \chi {\chi}^{\prime} )^{-1/2}.
\]

It only remains to show that 
equation (\ref{dfree}) does 
actually admit solutions. In 
Lorentzian and {\uh} 
signatures, we simply note that 
(\ref{dfree}) is automatically 
satisfied if we choose ${\alpha}_I$ 
and ${\alpha}_{I^{\prime}}$ 
to satisfy the Weyl equation
\be 
D^- {\alpha}_A = 0,\qquad
D^+ {\alpha}_{A^{\prime}} = 0.
\label{weyl} 
\ee 
On a general space of Lorentzian or 
{\uh} signature, equations 
(\ref{weyl}) will each have two 
linearly-independent solutions, and so we 
have a well defined new basis. In the 
Riemannian case, there is a vanishing 
theorem for solutions of the Weyl equation 
on compact manifolds with 
non-negative non-vanishing scalar curvature. 
However in this case we note that 
if we combine ${\alpha}_A$ and 
${\alpha}_{A^{\prime}}$ into a 
pair of Dirac spinors 
${\psi}_1 = {\alpha}_0 \oplus 
{\alpha}_{0^{\prime}}, 
{\psi}_2 = {\alpha}_1 \oplus 
{\alpha}_{1^{\prime}}$, 
then (\ref{dfree}) 
is automatically satisfied if ${\psi}_i$ are 
eigenspinors of the Dirac operator. The 
existence of independent solutions then 
follows from the general theory of 
elliptic operators.

\vskip .3cm

A brief note on the 
reality conditions for the spinor 
bases in each real signature 
is perhaps in order. If 
we are working with a Lorentzian 
metrics then, as explained in the previous section, 
we can assume the original spinor bases 
$\{ {\epsilon}_A \}, 
\{ {\epsilon}_{A^{\prime}} \}$ 
are related by complex conjugation, 
with ${\overline{{\epsilon}_A}} = 
{\epsilon}_{A^{\prime}}$. Similarly, we 
can take the solutions ${\alpha}_A, 
{\alpha}_{A^{\prime}}$ to be 
complex conjugates, 
in which case the fields 
${\chi}$ and ${\chi}^{\prime}$ 
are complex conjugates. 

\medskip

As explained in the Appendix, 
in the case of {\uh} signature, 
we can take the original spinor bases 
$\{ {\epsilon}_A \}, 
\{ {\epsilon}_{A^{\prime}} \}$, 
and the spinor fields ${\alpha}_A$ and 
${\alpha}_{A^{\prime}}$ to be 
real and independent. In this case, 
therefore, $\chi$ and ${\chi}^{\prime}$ 
are automatically real.

\medskip

In the Riemannian case, the 
quaternion maps $j_{\pm}$ on 
$\vpm$ can be assumed to act 
on spinor bases as
\[
j_-{{\epsilon}_A}
= {\delta}_{AB} {\epsilon}^B,\qquad
j_+ {\epsilon}_{A^{\prime}} 
= - {\delta}_{A^{\prime}B^{\prime}} 
{\epsilon}^{B^{\prime}}.
\]
Similarly, due to the SU$(2)$ nature 
of the connections on $\vpm$, we may 
assume that the spinor fields 
${\alpha}_A$ and ${\alpha}_{A^{\prime}}$ 
obey
\[
j_-{{\alpha}_A}
= {\delta}_{AB} {\alpha}^B,\qquad
j_+ {\alpha}_{A^{\prime}} 
= - {\delta}_{A^{\prime}B^{\prime}} 
{\alpha}^{B^{\prime}}.
\]
Due to the reality of the 
symplectic forms ${\epsilon}_{\pm}$, 
the functions $\chi$ and 
${\chi}^{\prime}$ are automatically real. 

\medskip

An important point to note, since we are 
about to consider conformal transformations, 
is that the function $f$ is real 
in each signature.

\newsection{Conformal transformations}
\label{sec:cfltrans}

Given a physical metric ${\bf g}$, 
and a basis ${\bf e}_i$ for $\TCM$, 
which obey the gauge condition (\ref{gauge}). 
We now define a new set of vector fields
\be
{\hat {\bf e}}_a = f {\bf e}_a,
\label{cflvec}
\ee
and metric ${\hat {\bf g}}$ satisfying
\be
{\hat {\bf g}} = f^{-2} {\bf g}.
\label{cfltrans}
\ee
In terms of the spinor decomposition of 
the metric, we take the symplectic forms 
to transform as
\[
{\hat {\epsilon}}_{\pm} = f^{-1} {\epsilon}_{\pm}.
\]
In order to preserve the normalisation 
of the spin-bases and to conform with 
(\ref{cflvec}), we take the bundles 
$\vpm$ to have conformal weight
$-\frac{1}{2}$ with the spin-bases transforming as
\bann
{\epsilon}_{A} = f^{-1/2} {\hat {\epsilon}}_{A}, 
\qquad
{\epsilon}_{A^{\prime}} = 
f^{-1/2} {\hat {\epsilon}}_{A^{\prime}},
\\
{\epsilon}^A = f^{1/2} {\hat {\epsilon}}^A, 
\qquad
{\epsilon}^{A^{\prime}} = 
f^{1/2} {\hat {\epsilon}}^{A^{\prime}}.
\eann

Using the symplectic forms ${\epsilon}_{\pm}$, 
we can view the curvature spinors ${}^- W, {\Phi}$ 
as sections of $S^4 (\vm)^*$ and 
$S^2 (\vm)^* \otimes S^2 (\vp)^*$ 
respectively. Defining 
\be
{\bf{\hat \Upsilon}} = 
{\hat d} (\log f) \in {\Gamma}(\form{1}),
\label{up}
\ee
we now recall that under the 
conformal transformation defined 
in equation (\ref{cfltrans}) the 
curvature spinors transform as 
follows \cite{PR,Be}:
\ba
{}^+ W &=& {}^+ {\hat W},
\\
{\Phi} &=& 
{\hat \Phi} - 2 \left( {\hat {\nabla}}{\bf{\hat \Upsilon}} - 
{\bf {\hat \Upsilon}} \otimes 
{\bf {\hat \Upsilon}} \right)
- 
\frac{1}{2} 
\left( {\hat\delta} {\bf{\hat \Upsilon}} + 
|{\bf{\hat \Upsilon}}|^2 \right) {\hat {\bf g}},
\\
f^2 s &=& {\hat s} 
+ 6 \, {\hat\delta}{\bf{\hat \Upsilon}} 
- 6 |{\bf{\hat \Upsilon}}|^2.
\ea

In order to simplify the analysis 
slightly, it is helpful to introduce 
a pair of arbitrary spinor fields 
$\pi \in {\Gamma}(\vm), \lambda 
\in {\Gamma}(\vp)$. We assume that 
these fields behave under conformal 
transformations as
\[ 
{\hat \pi} = \pi, \qquad 
{\hat \lambda} = \lambda.
\]
We then consider the scalar fields
\[
\begin{array}{ll}
\Psi \equiv
{}^+ W (\lambda, \lambda, \lambda, \lambda),
&{\hat \Psi} \equiv 
{}^+ {\hat W} ({\hat \lambda}, {\hat \lambda}, 
{\hat \lambda}, {\hat \lambda}),
\\
\Phi \equiv 
{\Phi} (\lambda, \lambda, \pi, \pi),
&{\hat \Phi} \equiv 
{\hat \Phi} ({\hat \lambda}, {\hat \lambda}, 
{\hat \pi}, {\hat \pi}).
\end{array}
\]
Since the spinor fields ${}^+ W$ and 
${\Phi}$ are totally symmetric in the 
relevant spinor indices we can recover all of the 
information of ${}^+ W$ and 
${\Phi}$ from the quantities $\Psi$ and $\Phi$ as the 
spinor fields $(\lambda, \pi)$ vary over ${\Gamma}({\Bbb
V}^{\pm})$.

We now wish to rewrite the curvature 
components of the physical metric 
${\bf g}$ in terms of the 
spin-connection of the unphysical 
metric ${\hat{\bf g}}$. To do this, we 
need to expand ${}^+ {\hat W}, {\hat \Phi}$ and 
${\hat s}$ in terms of the spin-connection 
${\hat {\gamma}}_{A^{\prime}}{}^{B^{\prime}}$. {}From equations 
(\ref{curvature}), (\ref{sdasdorth}), 
(\ref{curvdecomp}) and (\ref{ort}), 
we find that
\bann 
{\hat \Psi} &=&
{\hat \lambda}^{A^{\prime}} {\hat \lambda}^{B^{\prime}} 
{\hat \lambda}^{C^{\prime}} {\hat \lambda}^{D^{\prime}}
\left[
{\hat \nabla}_{AA^{\prime}}
{\hat \gamma}^A{}_{B^{\prime}C^{\prime}D^{\prime}} 
-
{\hat C}_{A^{\prime}B^{\prime}}{}^{EE^{\prime}}
{\hat \gamma}_{EE^{\prime}C^{\prime}D^{\prime}}
+ {\hat \gamma}_{AA^{\prime}B^{\prime}E^{\prime}}
{\hat \gamma}^A{}_{C^{\prime}D^{\prime}}{}^{E^{\prime}}
\right],
\\
{\hat \Phi} &=& 
2 {\hat \lambda}^{A^{\prime}} {\hat \lambda}^{B^{\prime}} 
{\hat \pi}^A {\hat \pi}^B
\left[
{\hat \nabla}_{AC^{\prime}}
{\hat \gamma}_B {}^{C^{\prime}}
{}_{A^{\prime}B^{\prime}} 
-
{\hat C}_{AB}
{}^{CC^{\prime}}
{\hat \gamma}_{CC^{\prime}A^{\prime}B^{\prime}}
+ 
{\hat \gamma}_{AI^{\prime}C^{\prime}B^{\prime}}
{\hat \gamma}^{I^{\prime}} {}_{BA^{\prime}}{}^{C^{\prime}}
\right],
\\
{\hat s} &=& - 4
\left[
{\hat \nabla}_{AA^{\prime}} 
{\hat \gamma}^A{}_{B^{\prime}} {}^{A^{\prime}B^{\prime}} 
-
{\hat C}_{A^{\prime}B^{\prime}}{}^{CC^{\prime}}
{\hat \gamma}_{CC^{\prime}}{}^{A^{\prime}B^{\prime}}
+{\hat \gamma}_{AA^{\prime}B^{\prime}}{}^{(A^{\prime}}
{\hat \gamma}_{C^{\prime}}{}^{|A|C^{\prime})B^{\prime}}
\right].
\eann

Our goal is now to use these 
equations to rewrite ${\Psi}, 
{\Phi}$ and $s$ 
in terms of the commutators of the 
vector fields ${\hat {\bf e}}_a$. We now 
define new fields
\[
{\hat \chi}_{AA^{\prime}} = 
{\hat C}_{AB}{}^{B}{}_{A^{\prime}},
\qquad
{\overline {\hat \chi}}_{A^{\prime}A} = 
{\hat C}_{A^{\prime}B^{\prime}}
{}^{B^{\prime}}{}_{A}.
\]
In terms of these fields we have
\[
{\hat \Gamma}_{a} \equiv {\hat C}_{ab}{}^b = 
- {\hat \chi}_a - {\overline {\hat \chi}}_a.
\]

Using the fact that
\[
{\hat \gamma}_{AA^{\prime}B^{\prime}C^{\prime}} = 
- {\hat C}_{A^{\prime}(B^{\prime}C^{\prime})A} 
+ \frac{1}{2}{\hat \Gamma}_{A(B^{\prime}} 
\epsilon_{C^{\prime})A^{\prime}},
\]
we find that
\setbox1=\vbox spread -5pt{\hsize 23pc 
$$f^2 {\Psi} = 
{\hat \lambda}^{A^{\prime}} {\hat \lambda}^{B^{\prime}} 
{\hat \lambda}^{C^{\prime}} {\hat \lambda}^{D^{\prime}}
\left[
{\hat \nabla}_{A^{\prime}}^A
{\hat C}_{B^{\prime}C^{\prime}D^{\prime}A} 
+ 2
{\hat C}^{E^{\prime}} {}_{A^{\prime}B^{\prime}}{}^E
{\hat C}_{C^{\prime}D^{\prime}E^{\prime}E}
\right]$$}
\setbox2=\vbox spread -5pt{\hsize 5.5in 
$$ f^2 {\Phi} = 
2 {\hat \lambda}^{A^{\prime}} 
{\hat \lambda}^{B^{\prime}} 
{\hat \pi}^A {\hat \pi}^B
\left[
{\hat \nabla}_{A}^{C^{\prime}}
{\hat C}_{C^{\prime}A^{\prime}B^{\prime}B} 
+
{\hat C}_{AB}
{}^{CC^{\prime}}
{\hat C}_{C^{\prime}A^{\prime}B^{\prime}C}
- {\epsilon}^{D^{\prime}E^{\prime}}
{\hat C}^{C^{\prime}}{}_{(A^{\prime}D^{\prime})A}
{\hat C}_{C^{\prime}(B^{\prime}E^{\prime})B}
\right]$$}
\setbox3=\vbox spread -3pt{\hsize 7cm 
$$ f^2 s = 
2 \left[ {\hat \nabla}_a {\overline {\hat \chi}}^a 
- {\hat C}_{A^{\prime}B^{\prime}C^{\prime}A} 
{\hat C}^{A^{\prime}C^{\prime}B^{\prime}A}\right]$$}
\ba
&\boxit{\box1}&
\label{psihat}
\\
&\boxit{\box2}&
\label{phihat}
\\
&\boxit{\box3}&
\label{lamhat}
\ea

We also find that, by construction, 
the commutator functions 
${\hat C}_{ab}{}^{c}$ obey the condition
\[
{\hat C}_{ab}{}^b = 
2 {\nabla}_a \log f.
\]
This condition is equivalent to the existence 
of ${\bf {\hat \omega}} \in \Gamma (\form{4})$ with 
the properties that
\setbox4=\vbox spread -7pt{\hsize 1.5in 
$${\cal L}_{{\bf {\hat e}}_a} {\bf {\hat \omega}} = 0$$}
\setbox5=\vbox spread -7pt{\hsize 2in 
$${\bf {\hat \omega}}({\bf {\hat e}}_{1}, {\bf {\hat e}}_2, 
{\bf {\hat e}}_3, {\bf {\hat e}}_4) = f^2$$}
\ba
&\boxit{\box4}&
\label{divfree}
\\
&\boxit{\box5}&
\label{volly}
\ea
where ${\cal L}$ denotes Lie 
derivative. In other words, 
the vector fields $\{ {\bf {\hat e}}_a \}$ are 
{\df} with respect 
to the volume element
${\bf {\hat \omega}}$.

\vskip .3cm
\noindent{\bf Remarks}
\vskip .3cm

The content of the vacuum Einstein equations 
is the vanishing of the expression for ${\Phi}$ 
and $s$ as given in equations (\ref{phihat}) 
and (\ref{lamhat}), and the {\df} 
condition (\ref{divfree}). There 
are a few properties of these equations 
that are of interest.

\vskip .2cm

\noindent 1. If ${\hat C}_{A^{\prime}B^{\prime}}{}^c = 0$, 
we automatically have that the physical 
metric is {\Rf} with anti-self-dual Weyl 
tensor \cite{MN}. More generally, if 
${\hat C}_{(A^{\prime}B^{\prime}C^{\prime})C} = 0$ then 
the Weyl tensor of both the physical 
and unphysical metric is anti-self-dual. 
Conversely, the vanishing of the self-dual 
Weyl tensor is equivalent to the 
condition that the totally symmetrised 
part of the spin-connection 
${\hat \gamma}_{(A^{\prime}B^{\prime}C^{\prime})C}$ be pure 
gauge, which in turn implies that 
${\hat C}_{(A^{\prime}B^{\prime}C^{\prime})C}$ be pure gauge. 

\vskip .2cm

\noindent 2. Only the expression for 
the trace-free Ricci tensor, equation 
(\ref{phihat}), has any explicit 
dependence on the unprimed coefficients 
${\hat C}_{AB} {}^{CC^{\prime}}$. 
It is geometrically reasonable that 
the Ricci tensor should depend on boths 
sides of the commutator: If ${\Phi} = 0$, the self-dual 
spin-connection ${\gamma}$ is 
an self-dual $\SLTC$ Yang-Mills 
field. As such, this connection 
will be integrable on anti-self-dual 
null planes. 
However, in a space with algebraically 
general anti-self-dual Weyl tensor, there are 
no integrable anti-self-dual null planes 
(in the above terminology, 
${\hat C}_{AB} {}^{CC^{\prime}} 
\neq 0$). Only if the anti-self-dual Weyl tensor 
vanishes can we fix a local frame for 
$\TCM$ with ${\hat C}_{AB} {}^{CC^{\prime}}$ 
vanishing within the present formalism. We 
then have our full quota of anti-self-dual null planes, 
and we are left with a problem that involves 
only the primed coefficients 
${\hat C}_{A^{\prime}B^{\prime}}{}^{CC^{\prime}}$.

\vskip .2cm

\noindent 3. Equations (\ref{psihat}), (\ref{phihat}), 
(\ref{lamhat}) and (\ref{divfree}) are 
accompanied by the Jacobi identity for 
the vector fields $\{ {\bf{\hat e}}_{i} \}$, 
which we write in the form 
\be
{\epsilon}^{abcd} 
\left[ {\bf{\hat e}}_b, 
\left[ {\bf{\hat e}}_c, {\bf{\hat e}}_d 
\right] \right] = 0. 
\label{jac}
\ee
We now note that we may rewrite (\ref{phihat}) 
and (\ref{lamhat}) in the form
\bann
f^2 \Phi &=& 
2 {\hat \lambda}^{A^{\prime}} {\hat \lambda}^{B^{\prime}} 
{\hat \lambda}^{C^{\prime}} {\hat \lambda}^{D^{\prime}}
\left[
\frac{1}{2}
< \left[ {\bf{\hat e}}_A^{C^{\prime}},
\left[ {\bf{\hat e}}_{A^{\prime}D}, 
{\bf{\hat e}}_{C^{\prime}}^D \right] \right], 
{\bf {\hat \epsilon}}_{BB^{\prime}} >
+
{\hat \chi}_{AB^{\prime}}
{\overline {\hat \chi}}_{A^{\prime}B}
+
{\hat C}_{AB}{}^{CC^{\prime}}
{\hat C}_{A^{\prime}B^{\prime}CC^{\prime}}
\right],
\\
f^2 s &=& 
- < \left[ {\bf{\hat e}}^{AA^{\prime}},
\left[ {\bf{\hat e}}_{BA^{\prime}}, 
{\bf{\hat e}}^{BB^{\prime}} \right] \right], 
{\bf {\hat \epsilon}}_{AB^{\prime}} >
+ 2 {\hat \chi}_c {\overline {\hat \chi}}^c.
\eann
Using (\ref{jac}), and translating all spinor 
indices back into tetrad indices, we find that 
the Einstein tensor of the physical metric 
${\bf g}$ obeys the relation
\[
f^2 \left(
r ({\bf v}, {\bf v}) - 
\frac{s}{2} {\bf g}({\bf v}, {\bf v}) \right) 
= 
v^{a} v^{b} 
\left[ 
{\hat \eta}^{cd}
< \left[ {\bf{\hat e}}_{c},
\left[ {\bf{\hat e}}_{d}, 
{\bf{\hat e}}_{a} \right] \right], 
{\bf{\hat \epsilon}}_{b} >
+
2{\hat \chi}_{a}{\overline {\hat \chi}}_{b}
-
{\hat \eta}_{ab}
{\hat \chi}_{c}{\overline {\hat \chi}}^{c}
-
2{}^+ \! {\hat C}_{a}{}^{cd}
\ {}^- \! {\hat C}_{bcd}
\right],
\]
where ${\bf v}$ is an arbitrary vector field, and
\[
{}^{+} \! {\hat C}_{ab}{}^{c} = 
\frac{1}{2} 
\left[
{\hat C}_{ab}{}^{c} + 
\frac{i}{2} {\epsilon}_{ab}{}^{de}
{\hat C}_{de}{}^{c} 
\right], \qquad
{}^- \! {\hat C}_{ab}{}^{c} = 
\frac{1}{2} 
\left[
{\hat C}_{ab}{}^{c} - 
\frac{i}{2} {\epsilon}_{ab}{}^{de}
{\hat C}_{de}{}^{c} 
\right],
\]
and
\[
{\hat \chi}_a = - {}^{+} \! {\hat C}_{ab}{}^{b},
\qquad
{\overline {\hat \chi}}_a = 
- {}^{-} \! {\hat C}_{ab}{}^{b}.
\]

The {\Rf}ness of the physical metric 
is then summarised in the equation
\setbox6=\vbox spread -3pt{\hsize 11cm 
$${\hat \eta}^{cd}
< \left[ {\bf{\hat e}}_{c},
\left[ {\bf{\hat e}}_{d}, 
{\bf{\hat e}}_{(a} \right] \right], 
{\bf {\hat \epsilon}}_{b)} > 
= 
- 2 {\hat \chi}_{(a}{\overline {\hat \chi}}_{b)} 
+
{\hat \eta}_{ab}
{\hat \chi}_{c}{\overline {\hat \chi}}^{c}
+
2{}^+ \! {\hat C}_{(a}{}^{cd}
\ {}^- \! {\hat C}_{b)cd}$$}
\be
\boxit{\box6}
\label{vaccy}
\ee
along with the {\df} condition 
(\ref{divfree}). The conformal factor $f$ 
which defines the transformation back 
to the physical metric is deduced 
from equation (\ref{volly}). 

Equation (\ref{vaccy}) is a 
form of the Einstein equations 
discussed in \cite{G}, where it
was noted that the object on 
the left-hand-side of the equation
can be interpreted as a 
generalisation of the
Yang-Mills operator for a 
constant connection on flat
space time, ${\Bbb M}$, 
taking values in the algebra of
{\df} vector fields on the 
auxiliary four-manifold $\M$.
The term on the right-hand-side 
of (\ref{vaccy}) is a source term,
which is purely an interaction 
between the {\sd} and {\asd} 
parts of the gravitational field. 
In the current context, this 
equation is interesting because 
it is symmetrical between the 
{\sd} and {\asd} parts of the
gravitational field, even though 
the formalism we started does not 
explicitly have such a symmetry. 
Whether a chiral or non-chiral 
approach to the 
Einstein equations is more 
generally useful probably 
remains to be seen, especially 
since most chiral formalisms 
are very sensitive to the 
dimension of the space-time 
we choose to work with. 

\newsection{Complex structures and torsion}
\label{sec:conn}

We have shown in the previous section how, 
starting from the self-dual $2$-form approach 
to the Einstein equations, we can partially 
fix spin-frames in a way that leads to a 
chiral description of the Einstein equations 
in terms of a frame of {\df} vector fields. 

Conversely, a solution of the vacuum Einstein 
equations in the the latter approach would 
correspond to a set of vector fields
${\bf {\hat e}}_i$ and a function $f$ which were 
solutions of equations (\ref{vaccy}), 
(\ref{divfree}) and (\ref{volly}). One can 
then return to the physical metric, and directly 
reconstruct the elements of the self-dual $2$-form 
approach, by simply reversing the conformal 
transformation given in equations 
(\ref{cflvec}) and (\ref{cfltrans}).
Although there is therefore a direct way of returning to
the self-dual variables, the connection with the {\df} vector
field approach does seem to suggest investigation of some
other aspects of the self-dual $2$-forms approach. 

In this section, we will concentrate on real 
Riemannian spaces, since this seems to be the 
context where the geometrical interpretation 
is most clear. We begin by restating the 
equations of the self-dual two-form 
formalism in this signature. 
We have a set of $2$-forms, ${\Sigma}_{A^{\prime}B^{\prime}}$, 
and an $\frak{su}(2)$ connection 
${\bf \gamma}$ which obey the equations 
\ba
&{\Sigma}_{A^{\prime}B^{\prime}} {\Sigma}^{C^{\prime}D^{\prime}} = 
{\epsilon}_{(A^{\prime}}{}^{C^{\prime}} 
{\epsilon}_{B^{\prime})}{}^{D^{\prime}} {\bf \nu},&
\label{cdj1}
\\
&d \Sigma + \left[ {\bf \gamma} , \Sigma \right] = 0,&
\label{cdj2}
\\
&R_{A^{\prime}B^{\prime}} = 
{}^+ W_{A^{\prime}B^{\prime}C^{\prime}D^{\prime}} 
{\Sigma}^{C^{\prime}D^{\prime}} + 
{\Phi}_{ABA^{\prime}B^{\prime}}
{\Sigma}^{AB} + 
\frac{s}{12} {\Sigma}_{A^{\prime}B^{\prime}}.&
\label{cdj3}
\ea
For the moment, we concentrate on finding the natural
analogues of the {\df} equation (\ref{divfree})
and the equation for the conformal factor (\ref{volly}) in
terms of the $2$-forms ${\Sigma}_{A^{\prime}B^{\prime}}$.

We begin by can defining the dual, ${\tilde \nu} \in \Gamma
({\wedge}^4 \TM)$, of the volume form ${\bf \nu}$ by the
condition 
\[
{\tilde \nu}
(\nu) = 1.
\]
The self-duality of the $2$-forms 
${\Sigma}_{A^{\prime}B^{\prime}}$ is expressed by the relation 
\[
{\tilde \nu} ({\Sigma}_{A^{\prime}B^{\prime}}) = 
{}^{\sharp} {\Sigma}_{A^{\prime}B^{\prime}}.
\]
We also find that
\[
{\tilde \nu}
({\bf \epsilon}^{AA^{\prime}} \wedge {\Sigma}^{B^{\prime}C^{\prime}}) = 
\, {\epsilon}^{A^{\prime}(B^{\prime}} 
{\bf e}^{C^{\prime})A}. \]

We now define the set of vector fields 
\be
{\bf v}^{A^{\prime}B^{\prime}} = 
{\tilde \nu}(*, d{\Sigma}^{A^{\prime}B^{\prime}}).
\label{vdef}
\ee
These vector fields are automatically 
{\df} with respect to ${\bf \nu}$ in the
sense that
\[
{\cal L}_{{\bf v}^{A^{\prime}B^{\prime}}} {\bf \nu} = 0.
\]
{}From the definition of ${\bf v}^{A^{\prime}B^{\prime}}$ 
along with (\ref{gamc}), it follows
that 
\[
v_{A^{\prime}B^{\prime}} = 
- C_{A^{\prime}B^{\prime}}{}^c {\bf e}_c +
{\Gamma}_{A(A^{\prime}} 
{\bf e}_{B^{\prime})} {}^A.
\]
Rewriting the right-hand-side of this 
relation in terms of the unphysical vector
fields $\{ {\bf {\hat e}}_i \}$, we find that
\ba
v_{A^{\prime}B^{\prime}} &=&
- f^{-2} {\hat C}_{A^{\prime}B^{\prime}}{}^c {\bf {\hat e}}_c
\nonumber\\
&=&
- \frac{1}{2} f^{-2} {\hat \epsilon}^{AB} 
\left[ {\bf {\hat e}}_{AA^{\prime}}, 
{\bf {\hat e}}_{BB^{\prime}} \right].
\label{vab}
\ea
Thus, the {\df} vector fields that naturally occur in the
self-dual $2$-form approach are, up to a factor, simply the
self-dual part of the commutator of the conformally
transformed vector fields. 

By reversing the conformal 
transformation (\ref{cfltrans}), we deduce
that
\[
{\bf \nu} = f^2 {\bf {\hat \omega}}.
\]
Therefore, given the function $f$, 
the $4$-form ${\bf {\hat \omega}}$ and the
self-dual part of the commutator 
$\left[ {\bf {\hat e}}_i, {\bf {\hat e}}_j
\right]$, we may directly reconstruct the
vectors $v_{A^{\prime}B^{\prime}}$ 
and the volume form ${\bf \nu}$. 
This in turn, via equation (\ref{vdef}), 
gives us $d {\bf \Sigma}_{A^{\prime}} 
{}^{B^{\prime}}$. {}From this, along with 
the fact that the 
${\bf \Sigma}_{A^{\prime}}{}^{B^{\prime}}$ 
are self-dual with respect to the volume
form ${\bf \nu}$, means that 
locally we can reconstruct
the ${\bf \Sigma}_{A^{\prime}}
{}^{B^{\prime}}$ up to ambiguity 
\[
{\bf \Sigma}_{A^{\prime}}{}^{B^{\prime}}\mapsto 
{\bf \Sigma}_{A^{\prime}}{}^{B^{\prime}} 
+ d {\bf\chi}_{A^{\prime}}{}^{B^{\prime}},
\]
where
\[
{}^* d {\bf\chi}_{A^{\prime}}{}^{B^{\prime}} 
= d {\bf\chi}_{A^{\prime}}{}^{B^{\prime}}.
\]

In the particular case when 
${\hat C}_{A^{\prime}B^{\prime}}{}^i = 0$, we see from 
equations (\ref{psihat}), (\ref{phihat}) 
and (\ref{lamhat}) that the physical 
metric is {\Rf} and has anti-self-dual Weyl 
tensor \cite{MN}. In this case, 
$v_{A^{\prime}B^{\prime}} = 0$ and so 
$d {\bf \Sigma}_{A^{\prime}} {}^{B^{\prime}} =0$. 
These equations, along with the orthogonality 
condition (\ref{ort}) completely determine, 
on a simply connected region, metrics 
with anti-self-dual Riemann tensor \cite{Pl1}.

The form of ${\bf v}_{A^{\prime}B^{\prime}}$ in equation 
(\ref{vab}) suggests that it is the obstruction 
to the integrability of self-dual null planes 
in the complexified tangent space. Since all such 
planes are integrable if and only if the Weyl 
tensor is {\asd}, this suggests that the vector 
fields ${\bf v}_{A^{\prime}B^{\prime}}$ should 
be related to the {\sd} part of the Weyl tensor. 

Suppose we consider simply connected, Riemannian 
spaces with a {\Rf}, anti-self-dual metric, 
then we may interpret the objects above in 
terms of {\hk} geometry. As such, we have a 
metric ${\bf g}$ which is Hermitian with 
respect to three integrable {\cs}s 
$({\bf I}, {\bf J}, {\bf K})$ on $\TM$ 
which obey the quaternion algebra: 
\ba
&{\bf I}^2 = {\bf J}^2 = {\bf K}^2 = - 1,&
\\
&{\bf I} {\bf J} = {\bf K}, \qquad
{\bf J} {\bf K} = {\bf I}, \qquad
{\bf K} {\bf I}= {\bf J},&
\label{quat}
\ea
which are covariantly constant:
\be
{\nabla} {\bf I} = {\nabla} {\bf J} 
= {\nabla} {\bf K} = 0.
\label{coconst}
\ee
If the {\cs}s are integrable, then 
$({\bf I}, {\bf J}, {\bf K})$ define a 
{\hc} structure, and the conformal 
structure of the metric ${\bf g}$ 
is {\asd}. The integrability 
condition for the covariant constancy 
of the {\cs}s then implies that the 
self-dual spin-connection is flat, and 
therefore that the metric ${\bf g}$ 
is {\Rf}.

More concretely, it is always possible 
to choose a null basis for $\TCM$ where 
the inverse metric is as in equation 
(\ref{NPmet}), and where 
$({\bf I}, {\bf J}, {\bf K})$ may be 
represented in matrix form as
\be
\left(\matrix
{i&0&0&0\cr
0&-i&0&0\cr
0&0&-i&0\cr
0&0&0&i\cr}\right)
\qquad
\left(\matrix
{0&0&1&0\cr
0&0&0&-1\cr
-1&0&0&0\cr
0&1&0&0\cr}\right)
\qquad
\left(\matrix
{0&0&i&0\cr
0&0&0&i\cr
i&0&0&0\cr
0&i&0&0\cr}\right).
\label{ijk}
\ee
In terms of the null basis we 
have used, the integrability 
conditions for these structures 
implies that there exist functions 
$a, b, c, d$ with 
\be
\left[ {\bf e}_1, {\bf e}_4 \right] 
= a {\bf e}_1 + b {\bf e}_4,\qquad 
\left[ {\bf e}_2, {\bf e}_3 \right] 
= c {\bf e}_2 + d {\bf e}_3,\qquad 
\left[ {\bf e}_1, {\bf e}_2 \right] + 
\left[ {\bf e}_3, {\bf e}_4 \right] = 
- d {\bf e}_1 + b {\bf e}_2 
+ a {\bf e}_3 - c {\bf e}_4.
\label{intcs}
\ee
If these conditions are satisfied, 
then the Weyl tensor is anti-self-dual. 
Demanding that the {\cs}s be covariantly 
constant requires satisfaction of the 
integrability condition that the 
structures commute with the 
Riemann curvature
\[
\left[ R({\bf X}, {\bf Y}), {\bf I} \right] =
\left[ R({\bf X}, {\bf Y}), {\bf J} \right] = 
\left[ R({\bf X}, {\bf Y}), {\bf K} \right] = 0, \qquad
\forall {\bf X}, {\bf Y} \in \Gamma(\TCM).
\]
Since, in four dimensions, 
$({\bf I}, {\bf J}, {\bf K})$ define 
(via the metric) a three-dimensional 
subspace of $\form{2}$, which we can 
define to be $\form{2+}$, this tells 
us that the self-dual part of the Riemann 
curvature of the metric ${\bf g}$ must 
vanish. As such, the metric ${\bf g}$ 
is {\Rf} and anti-self-dual.

Given the three {\acs}s 
$({\bf I}, {\bf J}, {\bf K})$, 
it is useful to notice that if 
$(a, b, c)$ are the components 
of a unit vector in ${\Bbb R}^3$, 
then the combination 
$a {\bf I} + b {\bf J} + c {\bf K}$ 
also defines an {\acs} on $\TM$, 
so a {\hk} manifold actually admits 
an $S^2$ worth of {\acs}s \cite{HKLR}. We view 
this $S^2$ as a complex projective 
line ${\bf P}_1$, which is itself 
constructed from two copies 
$U, U^{\prime}$ of the complex plane 
with coordinates $\zeta, {\zeta}^{\prime}$, 
related by $\zeta = ( {\zeta}^{\prime})^{-1}$ 
on $U \cap U^{\prime}$. We then combine 
the three {\acs}s into the single object
\[
{\bf J}_{\zeta} = 
\frac{1}{1 + {\zeta}{\overline \zeta}}
\left(
(1 - {\zeta}{\overline \zeta})I + 
({\zeta} + {\overline \zeta})J + 
i ({\zeta} - {\overline \zeta})K
\right).
\]
If ${\bf v}$ is a vector with 
${\bf I} {\bf v} = i {\bf v}$, 
then we define ${\bf w} = {\bf v} 
+ {\zeta} {\bf K} {\bf v}$ which 
automatically has the property 
that ${\bf J}_{\zeta} {\bf w} = 
i {\bf w}$ \cite{HKLR}. As such, if we 
assume $({\bf I}, {\bf J}, {\bf K})$ 
are represented as in equation 
(\ref{ijk}), then a local basis 
for the $(1,0)$ space of 
${\bf J}_{\zeta}$ is given by
\[
{\bf e}_1 + i {\zeta} {\bf e}_3, \qquad
{\bf e}_4 + i {\zeta} {\bf e}_2,
\]
and in a similar fashion we find 
that a basis for the $(0, 1)$ space 
is given by
\[
{\bf e}_2 + i {\overline\zeta} {\bf e}_4, \qquad
{\bf e}_3 + i {\overline\zeta} {\bf e}_1.
\]

If we ask that ${\bf J}_{\zeta}$ is 
integrable, $\forall \zeta \in {\Bbb C}$, 
we recover equations (\ref{intcs}). 
More precisely, if we define the 
Nijenhuis torsion tensor $N$ by
\[
4 N({\bf X}, {\bf Y}) = 
\left[ {\bf X}, {\bf Y} \right] 
+ J \left[ J {\bf X}, {\bf Y} \right] 
+ J \left[ {\bf X}, J {\bf Y} \right] 
- \left[ J {\bf X}, J {\bf Y} \right],
\]
for ${\bf X}, {\bf Y} \in T_x \M$, then on 
the $(1, 0)$ space of ${\bf J}_{\zeta}$ 
defined above we find that
\bann
\left( 1 + {\zeta}{\overline \zeta} \right)
N({\bf e}_1 + i {\zeta} {\bf e}_3, 
{\bf e}_4 + i {\zeta} {\bf e}_2) &=&
\left[
C_{14}{}^3
+ i \zeta \left( C_{12}{}^3 + C_{34}{}^3 \right) 
+ {\zeta}^2 C_{23}{}^3
\right.
\\
& &\hskip 1cm \left.
- i \zeta
\left(
C_{13}{}^1 
+ i \zeta \left( C_{12}{}^1 + C_{34}{}^1 \right) 
+ {\zeta}^2 C_{23}{}^1 \right) \right] 
\left( {\bf e}_3 + i {\overline\zeta} {\bf e}_1 \right)
\\
& &
\left( 
C_{14}{}^2
+ i \zeta \left( C_{12}{}^2 + C_{34}{}^2 \right) 
+ {\zeta}^2 C_{23}{}^2
\right.
\\
& &\hskip 1cm \left.
- i \zeta
\left(
C_{14}{}^4 
+ i \zeta \left( C_{12}{}^4 + C_{34}{}^4 \right) 
+ {\zeta}^2 C_{23}{}^4 \right) \right) 
\left( {\bf e}_2 + i {\overline\zeta} {\bf e}_4 \right).
\eann
The parts of $C_{ab}{}^c$ which occur 
in this expression are the totally 
symmetrised parts of the symmetrised coefficients 
$C_{(A^{\prime}B^{\prime}C^{\prime})C}$, 
which in turn correspond to the 
conformally invariant part of the 
vector fields ${\bf v}_{A^{\prime}B^{\prime}}$. 
As such, when the conformal structure is 
not anti-self-dual, the torsion of the 
{\acs} ${\bf J}_{\zeta}$ is directly 
related to the {\df} vector fields 
${\bf v}_{A^{\prime}B^{\prime}}$.

Treating the parameter $\zeta$ as a 
constant, the tensor ${\bf J}_{\zeta}$ 
is covariantly constant when we have a 
{\hk} structure. In the more general 
case we wish to consider, the obstruction 
to the covariant constancy of 
${\bf J}_{\zeta}$ is given by
\be
\left( \left[ {\nabla}_{\bf X}, 
{\nabla}_{\bf Y} \right] - 
 {\nabla}_{\left[ {\bf X}, 
{\bf Y} \right]}\right) {\bf J}_{\zeta} = 
\left[ R ({\bf X}, {\bf Y}), 
{\bf J}_{\zeta} \right].
\label{intein}
\ee
If the metric is Einstein, with ${\Phi} = 0$, 
the right-hand-side of this equation vanishes 
if take the bivector ${\bf X} \wedge {\bf Y}$ 
to be {\asd}. As such, the condition that the 
metric be Einstein becomes the condition that 
the Riemann tensor, viewed as a $2$-form with 
values in End$(T_x \M)$, commutes with the 
{\acs} ${\bf J}_{\zeta}$ when acting on any 
{\asd} bi-vector. If the Weyl tensor is {\sd}, 
then such an {\asd} bi-vector would define an 
integrable {\asd} null plane ${\bf \Sigma}$, 
with equation (\ref{intein}) holding for any 
${\bf X}, {\bf Y} \in T_x {\bf \Sigma}$. 
Since such planes would be integrable, 
we can interpret the vanishing of the 
left-hand-side of equation (\ref{intein}) 
in this case as the integrability condition 
for the tensor ${\bf J}_{\zeta}$ to be 
covariantly constant on the surface ${\bf \Sigma}$. 
In the general case when ${}^- W \neq 0$, however, 
equation (\ref{intein}) has no such interpretation 
as an integrability condition.

Perhaps the most compact way of 
viewing is in quaternionic notation. 
Generally, we translate objects with values 
in the vector bundle associated with the 
vector representation of SO$(4)$ into 
quaternion valued objects, and those 
taking values in the vector bundle 
corresponding to the adjoint representation 
of SU$(2)$ into objects with values in the imaginary 
quaternions. For example, given a standard 
orthonormal tetrad ${\bf \epsilon}_i$ for 
the Riemannian metric
\[
{\bf g} = {\Sigma}_{i = 0}^3 {\delta}_{ij} 
{\bf \epsilon}_i \otimes {\bf \epsilon}_j,
\]
we combine the $1$-forms ${\bf \epsilon}_i$ 
into the quaternion-valued $1$-form
\[
{\bf \epsilon} = 
{\bf \epsilon}_0 + i {\bf \epsilon}_1 
+ j {\bf \epsilon}_2 + k {\bf \epsilon}_3,
\]
with quaternionic conjugate
\[
{\overline{\bf \epsilon}} = 
{\bf \epsilon}_0 - i {\bf \epsilon}_1 
- j {\bf \epsilon}_2 - k {\bf \epsilon}_3.
\]
In terms of these forms we may write the metric as
\[
{\bf g} = \frac{1}{2} \left[ {\bf \epsilon} 
\otimes {\overline{\bf \epsilon}} 
+ {\overline{\bf \epsilon}} 
\otimes {\bf \epsilon} \right],
\]
which is invariant under the 
${\rm SU}(2) \times {\rm SU}(2) 
/ {\Bbb Z}_2$ action corresponding 
to the left or right multiplication 
of ${\bf \epsilon}$ by unit 
modulus quaternions. 
The form ${\bf \epsilon} \wedge 
{\overline{\bf \epsilon}}$ gives 
us an {\iqv}d self-dual $2$-form
\be
{\bf \epsilon} \wedge {\overline{\bf \epsilon}} = 
- 2 i \left[ {\bf \epsilon}_0 \wedge {\bf \epsilon}_1 
+ {\bf \epsilon}_2 \wedge {\bf \epsilon}_3 \right] 
- 2 j \left[ {\bf \epsilon}_0 \wedge {\bf \epsilon}_2 
+ {\bf \epsilon}_3 \wedge {\bf \epsilon}_1 \right] 
- 2 k \left[ {\bf \epsilon}_0 \wedge {\bf \epsilon}_3 
+ {\bf \epsilon}_1 \wedge {\bf \epsilon}_2 \right],
\label{sdforms}
\ee
whilst ${\overline{\bf \epsilon}} 
\wedge {\bf \epsilon}$ furnishes 
us with {\iqv}d anti-self-dual $2$-form,
\[
{\overline{\bf \epsilon}} \wedge {\bf \epsilon} = 
2 i \left[ {\bf \epsilon}_0 \wedge {\bf \epsilon}_1 
- {\bf \epsilon}_2 \wedge {\bf \epsilon}_3 \right] 
+ 2 j \left[ {\bf \epsilon}_0 \wedge {\bf \epsilon}_2 
- {\bf \epsilon}_3 \wedge {\bf \epsilon}_1 \right] 
+ 2 k \left[ {\bf \epsilon}_0 \wedge {\bf \epsilon}_3 
- {\bf \epsilon}_1 \wedge {\bf \epsilon}_2 \right].
\]
The complex structures $({\bf I}, {\bf J}, {\bf K})$ 
we considered earlier are modelled upon left 
multiplication by $(i, j, k)$ of the form 
${\bf \epsilon}$. The {\sd} forms appearing 
in equation (\ref{sdforms}) are the 
corresponding K{\"a}hler forms when the 
metric ${\bf g}$ is {\hk}.

We may now encode the information of the 
self-dual spin-connection into the 
imaginary-quaternion-valued $1$-form
\[
{\bf \Gamma} = 
\frac{i}{2} 
\left[ {\bf \Gamma}_{01} + {\bf \Gamma}_{23} \right]
+ \frac{j}{2} 
\left[ {\bf \Gamma}_{02} + {\bf \Gamma}_{31} \right]
+ \frac{k}{2} 
\left[ {\bf \Gamma}_{03} + {\bf \Gamma}_{12} \right],
\]
which has the property that
\be
d \left( {\bf \epsilon} \wedge 
{\overline{\bf \epsilon}} \right) = 
2 {\bf \Gamma} \wedge 
{\bf \epsilon} \wedge {\overline{\bf \epsilon}}.
\label{quatconn}
\ee
The curvature of this connection is the {\iqv}d $2$-form
\bann
{\bf R} &=& 
d {\bf \Gamma} + 
\frac{1}{2} \left[ {\bf \Gamma}, {\bf \Gamma} \right]
\\
&=& {}^+ W {\bf \lambda}^+
+ {\Phi} {\bf \lambda}^-
+ \frac{s}{12} {\bf \epsilon} 
\wedge {\overline{\bf \epsilon}},
\eann
where we have adopted the notation 
of equation (\ref{curvdecomp}), and 
${}^{\pm}{\bf \lambda}$ are orthonmormal 
bases for $\form{2\pm}$. 

The volume form ${\bf \nu}$ 
is deduced from the relation
\[
\left( {\bf \epsilon} 
\wedge {\overline{\bf \epsilon}} \right) 
\wedge 
\left( {\bf \epsilon} 
\wedge {\overline{\bf \epsilon}} \right) 
= - 24 {\bf \nu}.
\]
Given this volume form, we may 
define the {\iqv}d vector field 
\[
{\bf v} = {\bf{\tilde\nu}} 
( d \left( {\bf \epsilon} \wedge 
{\overline{\bf \epsilon}} \right) ),
\]
with the property that 
\[
{\cal L}_{\bf v} {\bf \nu} = 0.
\]
It follows from equation (\ref{quatconn}) 
that the vector field ${\bf v}$ carries 
a large amount of the information of the {\sd} 
spin-connection ${\bf \gamma}$. 
If the metric is Einstein, the two-form 
${\bf R}$ is {\sd} and therefore vanishes 
when evaluated on any {\asd} bivector. 
Therefore, the action of left multiplication 
by $( i, j, k)$ of the form ${\bf \epsilon}$ 
will be covariantly constant on {\asd} 
bivectors in the sense that the left hand side 
of equation (\ref{intein}) will vanish. Unfortunately, 
it does not seem possible to find any 
straightforward interpretation of this 
condition directly in terms of the vector 
fields ${\bf v}$. 

Similarly, whether there exists a concise 
geometrical interpretation of the Einstein 
equations in terms of the {\acs} 
${\bf J}_{\zeta}$ and its torsion which 
could serve as a useful alternative 
to the usual {\sd} two-forms approach 
remains to be seen. More correctly, 
we should look on the object 
${\bf J}_{\zeta}$ as the projection 
to $\TM$ of the horizontal part of 
the natural {\acs} on the projective 
spin-bundle ${\bf P} \vp$ \cite{AHS}. 
This {\acs} on ${\bf P} \vp$ is integrable 
if and only if the Weyl tensor on 
$\M$ is {\asd}. If we allow ourselves 
to consider the full spin-bundle ${\Bbb V}$, 
there do seem to be differential forms 
which capture some of the content of the 
Einstein equations \cite{MF,N}. 
Since approaches, however, are 
generally non-chiral, requiring 
information about both 
the natural {\acs} ${\bf J}^+$ 
on $\vp$, and the corresponding structure 
${\bf J}^-$ on $\vm$. As such, it would appear 
that there may some redundancy in these 
descriptions. The underlying geometry of 
these constructions also seems rather unclear.

\newsection{Conclusion}

The main objective of this paper is 
to give a more concrete derivation of 
the results of \cite{G} 
where the Einstein equations were 
reformulated in terms of a set of {\df} 
vector fields on an auxiliary four-manifold, 
in a way which seemed to make some connection 
with the Yang-Mills equations. We have 
demonstrated how these equations naturally 
arise, after a conformal transformation, if 
we choose spinor bases which satisfied the 
Dirac equation. Since our starting point was 
the self-dual two-form formalism the equations 
we arrive at are actually a chiral version 
of the {\em Lsdiff}$\M$ equations, which reduce 
to the form given in \cite{G} with the help 
of the Jacobi identity.

With the {\df} vector approach in mind, 
we then investigated some other aspects 
of the self-dual two-forms approach. In 
particular, there naturally arises a 
particular set of {\df} vector fields 
related to the self-dual two-forms. 
In the Riemannian sector, these vector fields 
may be interpreted in terms of the torsion of 
projection to the tangent space of the natural 
{\acs} on the projective spin-bundle of the space.
The natural statement 
of the Einstein condition in this context 
is simply that the {\acs} 
commutes with the Riemann curvature 
evaluated on an {\asd} bivector. 
Unless the {\asd} Weyl tensor vanishes, 
however, this condition on the {\acs} 
cannot be interpreted as an integrability 
condition. This is a similar dilemma to the 
one we face when we interpret the Einstein 
condition in terms of the self-dual spin-connection 
being an self-dual SU$(2)$ Yang-Mills field. 
In this case, the self-duality 
condition on the Yang-Mills 
connection may be interpreted as the 
integrability condition for the 
existence of particular holomorphic 
vector bundles over the 
projective spin-bundle of our space, 
but only if the space itself is 
self-dual \cite{AHS}. Alternatively, 
the self-dual Yang-Mills equations are only 
integrable on an self-dual background, the 
obstruction to their integrability being 
precisely the anti-self-dual part of the Weyl tensor.

It should be noted that all considerations have 
been local in nature. In the use of spinor 
techniques, we are implicitly assuming the 
existence of a spin structure. This does not 
seem to be a significant problem, however, 
since locally any manifold is spin and also 
a large proportion of the spinor analysis we 
use carries through with only a projective 
spin structure, to which there is no topological 
obstruction. A more important global problem is 
that we are implicitly assuming our metric may 
be described in terms of the non-vanishing 
linearly-independent set of vector fields 
$\{ {\bf e}_i \}$. The existence of 
such a set of vector fields 
is known to place restrictions on the cohomology 
of the underlying manifold \cite{MS}. In our case, 
we require four non-vanishing linearly-independent 
vector fields on a four-dimensional manifold, which 
implies that the tangent bundle of the manifold is 
trivial (implicitly implying that the manifold 
admits a spin-structure anyway). 
However, it is possible to develop a 
gauge invariant version of the formalism used here, 
where full SO$(1, 3)$ invariance is restored and the 
internal bases are no longer constrained by the {\df} 
condition \cite{ugly}. Although such a formalism could 
be used globally, it is essentially equivalent to the 
usual compacted spin-coefficient formalism \cite{PR}. 
It does, however, lead to a straight-forward derivation 
of the main points of the Light-Cone-Cut description of 
conformally Einstein spaces \cite{FKN}. 

Purely within the approach taken here, 
however, it appears that we have a 
description of the vacuum Einstein 
equations which emphasises the properties 
of self-dual and {\asd} null planes and their (lack of) 
integrability. Whether the approach can 
lead to any more concrete insight in analysing 
the Einstein equations in full generality, 
or perhaps in the case of algebraically-special 
metrics, is under investigation.

\nnewsection{Acknowledgments}

The author is grateful to thank Riccardo 
Capovilla for several illuminating 
discussions concerning the self-dual 
two-form approach to General Relativity, 
and to Andrew Chamblin for discussions 
on global aspects of four-dimensional 
Einstein manifolds. He would also like 
to thank the University of Newcastle 
for the Sir Wilfred Hall fellowship, 
under which this work was carried out. 

\nnewsection{Appendix: Reality Conditions}

\renewcommand{\theequation}
{A.\arabic{equation}}

Although the reality conditions for 
real manifolds with metrics of Lorentzian 
signature were discussed in 
Section~\ref{sec:notn}, 
for the sake of generality we here discuss 
the reality conditions for metrics of other 
real signatures in four dimensions. As such, 
we begin with a complex four-manifold $\M$ 
with a holomorphic metric ${\bf g}$. 
For each $x \in \M$, a frame for
$\TM$ defines an isomorphism $T_x \M \cong {\Bbb C}^4
\cong {\Bbb C}(2)$, with a vector ${\bf V} \in T_x \M$
being mapped to a $2 \times 2$ matrix via
\be
T_x \M \ni {\bf V} \mapsto 
{\tilde V} = \frac{1}{\sqrt{2}}
\left(\matrix
{V^0 + V^3&V^1- i V^2\cr
V^1+ i V^2&V^0 - V^3\cr
}\right) \in {\Bbb C}(2).
\label{tmisom}
\ee
We then have
\be
{\rm det} {\tilde V} = 
\frac{1}{2} {\bf g}({\bf V}, {\bf V}).
\label{metdet}
\ee
The map ${\tilde V} \mapsto L {\tilde V} R$, where
$L, R \in \SLTC$ gives an explicit description of the
isomorphism SO$(4, {\Bbb C}) \cong \SLTC \times
\SLTC /{\Bbb Z}_2$. Passing to the double cover, with $Spin(4,
{\Bbb C}) \cong \SLTC \times \SLTC$, we introduce the
two-dimensional complex vector bundles of spinors, $\vpm$,
associated with the separate $\SLTC$ factors. Being $\SLTC$
bundles, these bundles inherit symplectic forms, denoted
${\epsilon}_{\pm}$, which define isomorphisms $\vpm \cong
(\vpm)^*$. The form of the metric given in equation
(\ref{metdet}) then agrees with that of equation
(\ref{metdec}). The map (\ref{tmisom}) gives the
isomorphism between $T_x \M$ and $\vx{-}{x} \times
\vx{+}{x}$. The $2$-form ${\bf \Sigma}$ introduced in
Section~\ref{sec:notn} defines an isomorphism between
$( \form{2-} )_x$ and the adjoint representation space of
$\SLTC$, so we can choose a basis for $\vp$ where
${\bf \Sigma}$ is represented by a trace-free $2\times 2$
matrix of complex $2$-forms.

We wish to find the real structures satisfied by
${\bf \Sigma}$ and the bundles $\vpm$ which characterise 
real pseudo-Riemannian metric of signature $(0, 4)$ and
$(2, 2)$ which we will refer to as Riemannian and {\uh}
respectively. 

In the Riemannian case, 
${\rm Spin}(4) \cong {\rm SU}(2) \times {\rm SU}(2)$, so
the bundles $\vpm$, in addition to the symplectic forms mentioned
above, inherit Hermitian forms denoted
$< \ , \ >_{\pm}$, which define isomorphisms $\vpm
\cong ({\overline{\vpm}})^*$. Alternatively, since ${\rm
SU} (2) \cong {\rm Sp}(1)$, the bundles have a
quaternionic structure, with anti-linear
isomorphisms $j_{\pm}: \vpm \rightarrow \vpm$, 
such that $j_{\pm}^2 = -1$. We can then identify the Hermitian
structures in the form
\bann
< u , v >_- &=&
{\epsilon}_- (u, j v), \qquad u, v \in \vm,
\\
< u , v >_+ &=&
{\epsilon}_+ (j u, v), \qquad u, v \in \vp,
\eann
with
\[
{\epsilon}_{\pm} (j u, j v) = 
{\overline{{\epsilon}_{\pm} (u, v)}}
\]
in both cases. Similarly, the bundles 
$S^m \vm \otimes S^n \vp$ 
inherit a quaternionic structure when $m+n$
is an odd integer, and a real structure 
when $m+n$ is even.

In the case of an {\uh} metric, 
${\rm Spin}(2, 2) \cong {\rm SL}(2, {\Bbb R}) 
\times {\rm SL}(2, {\Bbb R})$ and the 
bundles $\vpm$ inherit real structures 
i.e. anti-linear maps ${\sigma}_{\pm}: \vpm \rightarrow \vpm$
with ${\sigma}_{\pm}^2 = 1$. Therefore, $\vpm$ arise
naturally as the complexification of the bundles of real spinors,
$\vx{\pm}{r}$, which are invariant under
the maps ${\sigma}_{\pm}$. The higher spinor bundles 
$S^m \vm \otimes S^n \vp$ inherit real structures, which
single out real spinors to be those which lie in $S^m
\vx{-}{r} \otimes S^n \vx{+}{r}$.

If we wish, we can choose local bases for $\vpm$ 
adapted to the particular real structures present in each
signature. We may always choose bases ${\epsilon}_A =
(o, {\iota})$ for $\vx{-}{x}$, $(o^{\prime},
{\iota}^{\prime})$ for $\vx{-}{x}$, which are orthonormal
with respect to the symplectic forms ${\epsilon}_{\pm}$ 
in the sense of equation (\ref{epdown}). In the {\uh} case,
we may choose these basis spinors to lie 
in the real spin-bundles $\vx{\pm}{r}$, so identifying
the real structures ${\sigma}_{\pm}$ with complex
conjugation, we have
\[
{\rm {\Uh}}:\qquad {\overline o} = o, \qquad
\qquad {\overline {\iota}} = {\iota}, \qquad
{\overline {o^{\prime}}} = o^{\prime}, \qquad
\qquad {\overline {{\iota}^{\prime}}} = {\iota}^{\prime}.
\]
In this basis, a spinor is real 
if its components are real. In 
particular the matrix-valued 
$2$-form ${\bf \Sigma}$ becomes 
a real trace-free matrix, 
the components of the Weyl 
spinor ${}^- W$ and the 
Ricci spinor ${\Phi}$ are real. 
Finally, the vector fields 
$\{ {\bf e}_i \}$ of equation 
(\ref{NPvecs}) are real, 
and span the real tangent space.

In the Riemannian case, taking components 
of spinors in $\vx{\pm}{x}$ with respect 
to the bases defines isomorphisms 
$\vx{\pm}{x} \cong {\Bbb C}^2 \cong {\Bbb H}$. 
Given the form of the map (\ref{tmisom}), and 
the action of $\SLTC \times SLTC$ as 
${\tilde V} \mapsto L {\tilde V} R$, 
we represent elements of $\vx{+}{x}$ 
as row vectors, transforming as 
$\vx{+}{x} \in \pi \mapsto \pi R$ 
and elements of $\vx{-}{x}$ 
as column vectors, transforming as 
$\vx{-}{x} \in \psi \mapsto L \psi$. 
Given the components $(z_1, z_2)$ of 
an element of $\vx{+}{x}$, we map this 
to $w = z_1 + z_2 j \in {\Bbb H}$. 
The action of $R$ now corresponds to 
right multiplication by a unit 
modulus quaternion. 
The Sp$(1)$ property means that this 
action is compatible with 
left multiplication of $w$ by the 
unit quaternion $j$, which acts as 
$(z_1, z_2) \mapsto 
(- {\overline{z_1}}, {\overline{z_2}})$. 
This is the model for the quaternionic 
structure $j_+$ on $\vx{+}{x}$. 
Similarly, the action of the quaternionic 
structure $j_-$ on $\vx{-}{x}$ corresponds 
to right multiplication by $- j$, which is 
compatible with the action $L$ on 
$\vx{-}{x}$, which corresponds to 
left multiplication by a unit quaternion $l$.

In terms of the bases $(o, {\iota})$, 
and $(o^{\prime}, {\iota}^{\prime})$
we may therefore explicitly 
take $j_{\pm}$ to act as 
\bann
j_-(o) = {\iota}, \qquad j_- ({\iota}) = - o,
\\
j_+(o^{\prime}) = - {\iota}^{\prime}, \qquad 
j_+ ({\iota}^{\prime}) = o^{\prime},
\eann
or, more compactly
\[
{\rm Riemannian}:\qquad 
j_-{{\epsilon}_A}
= {\delta}_{AB} {\epsilon}^B,\qquad
j_+ {\epsilon}_{A^{\prime}} 
= - {\delta}_{A^{\prime}B^{\prime}} 
{\epsilon}^{B^{\prime}}.
\]
Identifying the real structures that occur on higher
bundles may be identified with 
complex conjugation, with respect 
to these spinor bases, the object ${\bf \Sigma}$ becomes a
trace-free skew-Hermitian matrix of two-forms.
The self-dual Weyl spinor ${}^+ W$, as a real
element of $S^4 \vp$, has components which
satisfy
\[ 
{\overline{{}^+ W_{A^{\prime}B^{\prime}C^{\prime}D^{\prime}}}} = 
{\delta}_{A^{\prime}E^{\prime}} {\delta}_{B^{\prime}F^{\prime}} 
{\delta}_{C^{\prime}G^{\prime}} {\delta}_{D^{\prime}H^{\prime}} 
{}^+ W^{E^{\prime}F^{\prime}G^{\prime}H^{\prime}}.
\]
Therefore, if the Weyl spinor is non-vanishing, 
the roots of the quartic polynomial 
${}^+ W ({\pi}, {\pi}, {\pi}, {\pi})$, 
where $\pi \in \vp$, occur in pairs of 
the form $\zeta = (\lambda, 
- 1 / {\overline{\lambda}})$, where 
$\zeta = \pi_{0^{\prime}} / \pi_{1^{\prime}}$ 
is an affine coordinate on ${\bf P}_1$. 
Therefore, in terms of the standard 
classification of Lorentzian Weyl 
tensors \cite{PR}, the self-dual Weyl 
tensor of a Riemannian metric is either 
algebraically general, type--$D$ or 
flat \cite{karl}. Finally, the complex 
vector fields $\{ {\bf e}_i \}$ of 
equation (\ref{NPvecs}) obey the 
reality conditions: 
\[
{\rm Riemannian}:\qquad
{\overline{{\bf e}_1}} = - {\bf e}_2,
\qquad{\overline{{\bf e}_2}} = - {\bf e}_1,
\qquad{\overline{{\bf e}_3}} = {\bf e}_4,
\qquad{\overline{{\bf e}_4}} = {\bf e}_3,
\]
and the real tangent space takes the form
\be
{\rm Riemannian}:
\qquad T_x \M = {\rm Span}_{\Bbb R} 
\left( 
i \left( {\bf e}_1 + {\bf e}_2 \right), 
{\bf e}_1 - {\bf e}_2, 
i \left( {\bf e}_3 - {\bf e}_4 \right), 
{\bf e}_3 + {\bf e}_4 
\right).
\ee

\end{document}